\documentclass[oneside,a4paper]{article}

\usepackage{hyperref}
\hypersetup{colorlinks,allcolors=black}

\usepackage{amsmath}
\usepackage{csquotes}
\usepackage{authblk}
\usepackage{multirow, makecell, amssymb}
\usepackage{graphicx}

\title{\textbf{Skyline Operators and Regret Minimization Techniques for Managing User Preferences in the Query Process}}
\author{Giulio Talarico}
\affil{Politecnico di Milano\\
Milan, Italy\\
\href{mailto:giulio.talarico@mail.polimi.it}{giulio.talarico@mail.polimi.it} }
\date{}

\begin{document}

\maketitle

\begin{abstract}

The problem of selecting the most representative tuples from a dataset has led to the development of powerful tools, among which Skyline and Ranking (or Top-$k$) queries stand out for their ability to support the optimization of multiple criteria in the query process. This paper surveys the remarkable efforts made towards the extension of the aforementioned tools to overcome their limitations, respectively the explosion of the output result and the difficulty of query formulation. Moreover, we explore the application of these state-of-the-art techniques as preference-based query frameworks, proposing a comparison of their query personalization capabilities, the ability to control the output size and their flexibility with respect to the user input preferences.

\end{abstract}


\section{Introduction}

During the last twenty years, since the introduction of the Skyline operator \cite{DBLP:conf/icde/BorzsonyiKS01}, numerous powerful tools have been proposed in the realm of database systems with the aim of assisting multi-criteria decision making, namely to provide the user with the most useful results with respect to a set of input preferences. 

Managing user preferences in the query process has been proved to be fundamental when dealing with large scale databases, where the user can get lost in a \textit{mare magnum} of potentially interesting data. Some representative scenarios where user-centered information retrieval is indeed requisite are: \textit{(i)} recommender systems \cite{DBLP:journals/aim/PuC08}, where user input preferences, extrapolated for instance from clickstream data, are used to restrict the output of proposed products only to those of significant value with respect to the user profile; \textit{(ii)} UX design, where both quantity and quality of the elements which the user can interact with are key aspects which influence the ease and the speed of the user's decision making \cite{doi:10.1080/17470215208416600}; \textit{(iii)} product search \cite{DBLP:conf/edbt/Chomicki02, 10.1145/335191.335423}, where the output result has to take into account a set of search criteria that not always the user is able to provide in a complete, consistent and quantitative manner, considering the investigative nature of a search with respect to unfamiliar domains.

In such complex scenarios, traditional approaches to multi-objective optimization, among which we refer to Skyline and Ranking (or Top-$k$) queries, suffer respectively from poor personalization capabilities, because of the \textit{ceteris paribus} semantics, and the lack of flexibility with respect to user input preferences, because of the need of a scoring function as a mean of defining user preferences; furthermore, both approaches are doomed by the curse of dimensionality, that makes the output of a skyline query often too big to handle (with the risk of paralyzing the decision making process of the user) and the weights used to compute the score of a high-dimensional tuple too hard to define in the first place. The aim of this survey is to compare different techniques developed to overcome such limitations, focusing on their ability of satisfying three hard requirements for \textit{user preference management} \cite{DBLP:conf/sigmod/MouratidisL021}: \textit{(1)} personalization, \textit{(2)} control over the output size, \textit{(3)} flexibility of the specified preferences.

In section 2 we summarize the state-of-the-art of tools and methodologies that improved the capabilities of traditional Skyline and Ranking queries, namely Flexible Skylines, Skyline Ranking and Regret Minimization queries. In section 3 we explore the application of these tools as frameworks for handling user preferences in the query process. Finally, in section 4, we briefly review and discuss the big picture of multi-objective query optimization approaches depicted in this survey.

\section{State of the art}

Despite their ability to capture the most interesting objects of a dataset with respect to a set of user preferences, the Skyline operator and Ranking queries both present some limitations: 

\begin{itemize}
    \item the core concept behind Skyline queries is the \textit{Pareto improvement principle}, which is the reason behind the simplicity of the Skyline semantics: the user is only asked to state his absolute preferences about each individual attribute without taking into account its relative importance with respect to the other attributes of the examined schema. In fact, Skyline queries will output the set of \textit{non-dominated} tuples: a tuple $p$ is said to \textit{dominate} a tuple $p'$ when $p$ is nowhere worse and better at least in one dimension (no matter which) than $p'$. In principle, this approach provides considerable summarization power, since the output of the Skyline query will be the set of \textit{potentially} interesting tuples, although there are some major drawbacks not only because all attributes are considered equally important, but also because it is more unlikely for a tuple to be dominated as the number of dimensions increases, hence the size of the Skyline will become too big to bring the user useful results: experimental evidence in \cite{DBLP:conf/vldb/GodfreySG05} shows that the output size of a skyline query can be exponential in the number of attributes; 
     
    \item in order to apply ranking algorithms \cite{DBLP:conf/pods/FaginLN01} so that only a predetermined number of relevant results is evaluated, the user is required to provide a \textit{scoring function} which represents his query preferences: this is typically expressed in the form of an aggregation function of partial scores defined for each attribute, thus allowing one to give different attributes different importance and, since a global comparison criterion is established, to select the desired number of tuples among the most relevant ones. On the other hand, specifying a scoring function is not always straightforward and it makes the \textit{preference elicitation} process much more complex since the user typically has to assign a numerical weight to each attribute; besides, small variations of the weights may lead to significant variations of the query results, therefore reducing the level of confidence of its relevance.
\end{itemize}

In the following sections we summarize, to the best of our knowledge, the main ideas behind some of the methods developed to combine the most effective characteristics of the aforementioned techniques, namely the simplicity of formulation and the finer control both over the output size and over the importance contribution of each attribute in the query process. For the purpose of this survey, three main categories are identified: Flexible Skylines, Skyline Ranking and Regret Minimization Queries.

\subsection{Flexible Skylines}

The flexibility introduced by this category of techniques comes from the fact that the user is not required to formulate a detailed \textit{scoring function}: instead, different approaches are embraced to integrate user preferences in a more general, but still representative way, into the Skyline framework, providing broader control over the query constraints, such as the possibility of expressing relative importance between attributes, introducing qualitative trade-offs, taking into account inaccuracies in the process of preference formulation and, accordingly, also reducing the query output size.

\subsubsection{R-Skylines}

Restricted Skylines \cite{DBLP:journals/pvldb/CiacciaM17} combine the notion of dominance and the notion of ranking introducing the concept of $\mathcal{F}$\textit{-dominance}: given a set of monotone scoring functions $\mathcal{F}$, a tuple t $\mathcal{F}$\textit{-dominates} another tuple $s \neq t$, denoted by $t\prec_{\mathcal{F}}s$, \textit{iff} $\forall f \in \mathcal{F} .\ f(t) \le f(s)$. Defining a family of scoring functions allows the user to specify a set of constraints over the weight space rather than explicitly set exact weight values for each attribute, as it is done instead in ranking queries: for instance in a 2-dimensional schema with attribute $A_1$ being \textit{price} and attribute $A_2$ being \textit{mileage} we may express a set of constraintc $C=\{w_1 \ge w_2\}$ for a family $\mathcal{F}$ of \textit{e.g.} linear scoring functions to express the preference of price over mileage. 

The R-Skyline coincides with the traditional skyline if $\mathcal{F}$ is the family of all the monotone scoring functions, whereas, as it is further analyzed in \cite{DBLP:journals/tods/CiacciaM20}, when finite sets of scoring functions are taken into account, R-Skylines incorporate the characteristics of \textit{dynamic skylines} \cite{DBLP:journals/tods/PapadiasTFS05}. The unique query potential of R-Skylines stands out when infinite sets of scoring functions are considered: in fact, experimental results demonstrate a considerable reduction of the query output size with respect to the result obtained using the traditional Skyline operator, suggesting that there is some correlation between the increase of the \textit{dominance} region, caused by the introduction of constraints over the weight space, and the reduction of output tuples, where the \textit{dominance} region of a tuple $p$ is defined as the set of all points dominated by $p$.

\subsubsection{Uncertain Top-k Queries}

A similar concept of flexibility is used in \cite{DBLP:journals/pvldb/MouratidisT18} to define Uncertain Top-k Queries (UTK): the weight vector (directly provided by the user or computationally inferred) is extended to a region in order to take into consideration the inherent uncertainty of the weight values provided by the user, hence computing results also for query preferences (weight vectors) that are similar (close) to the user defined preference profile (\textit{seed} vector); this is implemented in the definition of \textit{r-dominance}: given a region $R$ in the preference domain and a linear scoring function $S$, a tuple $p$ \textit{r}-dominates another tuple $p'$ when $S(p) \geq S(p')$ for any weight vector in $R$, and there is at least a weight vector in $R$ for which $S(p) > S(p')$. Specifying a region $R$ introduces an extra measure to break ties between incomparable tuples according to the traditional notion of dominance, since it suffices to check whether the vertices of $R$ are all inside or all outside the half-space $S(p) \geq S(p')$, which determines a region where $p$ \textit{r}-dominates $p'$ and another one where $p$ is \textit{r}-dominated instead. UTK result  contains tuples that are dominated by fewer than $k$ others, \textit{i.e.} the \textit{r}-skyband of the dataset.

UTK capabilities are further extended in \cite{DBLP:conf/sigmod/MouratidisL021} introducing the concept of \textit{$\rho$-dominance}, that only considers vectors that are within a distance $\rho$ from the \textit{seed}. This particular technique allows the user to specify not only a dominance parameter $k$, but also the desired output size $m$, which is used to tune the distance parameter $\rho$ such that it is the minimum distance that produces exactly $m$ tuples as a result, hence the closest to the user's preference profile. It is worth noting that when $\rho=0$ this is equivalent to performing a \textit{top-k} query, whereas when $\rho=\infty$ the output will be a \textit{k-skyband}.

\subsubsection{P-Skylines}

Prioritized Skylines \cite{DBLP:journals/vldb/MindolinC11} allow one to express relative importance between attributes through the use of \textit{p-expressions}: with this kind of approach user preferences are handled using a more complex preference framework that makes use of \textit{preference formulas}, introduced in \cite{DBLP:journals/tods/Chomicki03}. A \textit{p-expression} $\pi$ is the composition of attribute preference relations, used to express an absolute preference over the values of a single attribute (\textit{e.g.} price: the lower the better), with \& (\textit{prioritized accumulation}) and $\otimes$ (\textit{Pareto accumulation}) operators, used to define the relative importance of attribute preferences: $\pi_1\ \&\ \pi_2$ means that $\pi_1$ has higher importance than $\pi_2$, whereas $\pi_1\ \otimes\ \pi_2$ means that the two operands have the same importance. The composition of such relations is represented as a hierarchical structure using a \textit{p-graph}, which is then used to test the dominance of a tuple with respect to another one, hence to compute the most preferred tuple in a set.

In order to elicit user preferences and construct the corresponding \textit{p-expressions}, the authors propose a feedback-based approach that identifies superior and inferior examples, \textit{i.e.} desirable and undesirable tuples, bypassing the pairwise comparison of all attributes; this provides even more flexibility to the query process, since the user may be directly or indirectly involved: for instance, in a web centered scenario, one could take advantage of profiling information such as click-through rate to determine superior examples.

A similar technique, where user-specified qualitative partial preferences in the form of a graph guide the query process, is \texttt{Telescope} \cite{DBLP:conf/dasfaa/LeeYH07}.

\subsubsection{Trade-off Skylines}

Trade-off Skylines \cite{DBLP:conf/edbt/LofiGB10} deal with the introduction of flexibility in the query process integrating user preferences in the form of qualitative trade-offs: the idea is that a user may be willing to sacrifice the goodness of some attribute in favor of the improvement of some other attribute (\textit{e.g.} willingness to pay up to 5\$ more on a pizza order to get extra cheese, sacrificing price in favor of tastiness). Compensation semantics are formalized in \cite{DBLP:journals/ijcsa/BalkeGL07} through the notion of \textit{amalgamated preferences} and \textit{amalgamated equivalences}: the user can extend the Pareto semantics with additional domination relationships in order to obtain the so called \textit{generalized} skyline.

This enhancement brings to light some new difficulties: the additional trade-off semantics makes the dominance check among tuples more complex since the amalgamation of attribute domains breaks the property of separability of traditional skylines, which normally allows for a simple attribute-based comparison as dominance check criterion, thus the authors provide a tree-based algorithm to represent trade-offs and optimize the dominance check process, so that compromises can be efficiently taken into account in the skyline query process.

\subsection{Skyline Ranking}

The common objective of Skyline Ranking techniques is to select from the points of a skyline only the $k$ most useful ones, where the output size constraint $k$ is defined by the user. This is not trivial, since, by definition, all the points in a skyline are equally important and, since during the skyline query process the user is not required to provide any preference function that ranks each tuple accordingly, different criteria may be used to select exactly $k$ objects as a result; the following techniques will take into account structural characteristics of the skyline set to rank its items.

\subsubsection{Skyline Ordering}

The Skyline Ordering \cite{DBLP:journals/tkde/LuJZ11} technique focuses on the possibility of specifying arbitrary query size constraints, \textit{i.e.} a query output size parameter $k$ that is not known \textit{a priori} to be smaller or larger than the actual size $s$ of the skyline. This is achieved by combining \textit{pointwise ranking} and \textit{set-wide maximization}, two popular approaches used to reduce the size of the skyline: the former defines a total ordering among the skyline points using a scoring function, although it is not really suitable in the context of skyline queries since it is not only hard to formulate because of the sensitivity of the query results with respect to the function weights, but also tends to select points with similar features, since they will have similar scores, thus limiting the variety of the query results; the latter determines a subset of $k$ interesting data points by maximizing a collective property of the set, such as the total number of distinct points dominated by the points in the subset, although it is constrained to an output parameter $k < s$.

The algorithm to compute the \textit{skyline order} $S = \{S_1,S_2,...,S_n\}$ proceeds by iteratively computing each $S_i$, the so-called \textit{skyline subset}, as the skyline of $P \setminus \bigcup_{j=1}^{i-1} S_j$, where $P$ is the set of all points and $S_1$ is the skyline of $P$, until all points belong to a subset: thus, points in $P$ are partially ranked and partitioned in skyline layers that are then used to select the \textit{top-k} points; in fact each \textit{skyline subset} is sequentially checked and merged into the output set $S$ if its cardinality is less than the requested size $k$, until the remaining points have to be selected from the last skyline subset using a \textit{set-wide maximization} technique. 

\subsubsection{k-Representative Skyline}

The idea behind k-Representative Skyline is to quantify the descriptive power of points in a skyline set so that the user is able to make accurate decisions even though only a small sample of $k$ representative points of the dataset have been analyzed. The approach described in \cite{DBLP:conf/icde/LinYZZ07} builds up a definition of representativeness based on the dominance properties of the skyline points: in particular, $k$ points are selected such that the number of points dominated by one of the points in the set is maximized.

The same concept is proposed in \cite{DBLP:conf/icde/TaoDLP09}, although with a different definition of representativeness in order to capture more efficiently and more concisely the contour of the skyline: $k$ points are selected such that the maximum distance between a non-representative point and its closest representative is minimized. Experimental results show that the latter approach provides a more adequate summary of the dataset, with better precision as the parameter $k$ increases.

\subsubsection{SKYRANK}

The SKYRANK framework \cite{DBLP:journals/dke/VlachouV10} exploits the idea of link-based ranking to select the most interesting tuples from a skyline set. In order to do that, a graph-based structure is built, called \textit{skyline graph}, where nodes are all the skyline points and edges are the \textit{subspace dominance relationships}, which are used to determine the interestingness of the skyline points: given a dataset $D$ a point $p \in D$ is said to \textit{dominate} another point $q \in D$ on subspace $U$, denoted as $p \prec_q q$ if (1) on every dimension $d_i \in U$, $p_i \le q_i$ and (2) on at least one dimension $d_j \in U$, $p_j < q_j$. This approach integrates both the concept of \textit{k-dominance} \cite{DBLP:conf/sigmod/ChanJTTZ06}, which is introduced to increase the probability of a point to be dominated by simply relaxing the notion of dominance to a subset of dimensions, thus decreasing the skyline size, and the concept of \textit{skyline frequency} \cite{DBLP:conf/edbt/ChanJTTZ06}, which takes into account how often tuples are returned as a result of a skyline query when different dimensions are considered.

By construction, the SKYRANK framework also supports personalized Top-k Skyline queries, since it is able to take into account user preferences in the form of a particular subset of dimensions $PRF=\{U_1, ..., U_m\}$ used to compute the \textit{subspace dominance relationships} between skyline points, where each dimension represent a specific attribute of interest of the user's query.

\subsection{Regret Minimization Queries}

This category of techniques approaches the problem of multi-criteria decision making with a different perspective: an unknown scoring function is assumed, as in the traditional Skyline scenario, therefore, in order to select the most interesting tuples from the skyline set a \textit{regret} measure is taken into account, \textit{i.e.} the disappointment of the user seeing the k-regret result set with respect to the result he would have expected from the whole dataset. A thorough survey on regret minimization queries is proposed in \cite{DBLP:journals/vldb/XieWL20}, here we summarize some of the most interesting techniques.

\subsubsection{k-Representative Regret Minimization}
The \textit{k-regret} operator \cite{DBLP:journals/pvldb/NanongkaiSLLX10} computes a subset of $k$ skyline points by minimizing the so called \textit{maximum regret ratio}: given a dataset $D$, a subset of points $S \subseteq D$ and a scoring function $f$, the \textit{regret ratio} is computed as \[rr_D(S,f) = \frac{gain(D,f)-gain(S,f)}{gain(D,f)}\] where the \textit{gain} of a set is the maximum score obtained through the specified utility function $f$ by some point belonging to that set; in this particular scenario though, the scoring function $f$ is not known, hence the \textit{maximum regret ratio} is introduced, which takes into account a class of scoring functions $\mathcal{F}$. In particular, the authors demonstrate that for non-decreasing linear scoring functions, which in practice are good models of human reasoning for preference assignment, both a lower and an upper bound can be found as a function of the output parameter $k$ and the dimensionality $d$: this makes the \textit{k-regret} operator not only \textit{scale-invariant} but also \textit{stable}, \textit{i.e.} adding unimportant tuples in the dataset will not change the output result. 

The concept of \textit{regret minimization} here described focuses on the minimization of the user's regret while guaranteeing the query output size to be $k$, although there exist some other studies \cite{DBLP:conf/icde/XieW0T20, DBLP:conf/wea/AgarwalKSS17} which take on the same problem focusing on the minimization of the output size while guaranteeing a specified happiness ratio $\alpha$.

\subsubsection{k-Regret Minimizing Set}

The authors of \cite{DBLP:journals/pvldb/ChesterTVW14} base their research on the \textit{k-regret operator} proposed in \cite{DBLP:journals/pvldb/NanongkaiSLLX10}: in particular they relax the criterion for determining user happiness by generalizing the \textit{regret ratio}, which considers the theoretical best point of a subset, to a \textit{k-regret ratio} which measures how far from the $k$-th best point is the best point; thus, given a dataset $D$, a subset of points $S \subseteq D$ and a weight vector $w$, the \textit{k-regret ratio} is computed as \[k\text{-}regratio(S,w) = \frac{max(0, kgain(D,w) - 1gain(S,w))}{kgain(D,w)}\] where the \textit{kgain} represents the score of the $k$-ranked point in $S$, hence 1\textit{gain} corresponds to the definition of \textit{gain} previously introduced in \cite{DBLP:journals/pvldb/NanongkaiSLLX10}. The \textit{maximum k-regret ratio} is computed taking into account all possible weight vectors $w$. This new \textit{regret} definition does not require to include in the query result the user's top choice in order to make him happy. Since $k$ is used as a rank parameter, the parameter that constrains the output size is now $r$, and an algorithm is presented to compute the \textit{k-regret} minimizing set of order $r$ both for 2 dimensional datasets and, even though not optimal, for more than 2 dimensions.

An interesting extension of the \textit{k-regret} operator is introduced in \cite{DBLP:conf/IEEEwisa/DongZQH18}, where an additional set of binary constraints provided by the user is taken into consideration: this particular technique aims at selecting points that not only minimize the \textit{maximum regret ratio}, but also maximize the satisfaction of a set of binary constraints $C = \{c_1,...,c_r\}$, where a generic $c_i$ is a \textit{true/false} condition.

\subsubsection{ReDi}

The purpose of ReDi \cite{DBLP:conf/sigmod/HussainKS15} is to combine the principle of regret minimization with the one of diversity maximization; these two approaches share the objective of identifying a representative set of tuples among the full dataset, even though they serve very different needs: the former aims at determining a subset of data that satisfies the user the most, the latter at providing data that has good coverage on the dataset and low redundancy. Nevertheless, as the authors point out, typically the user is able to provide query preferences only for a subset of attributes, expecting to get some more insights about the attributes that have not been taken into consideration from the result itself: hence the need of capturing diverse objects that summarize the unsolicited part of information that is stored in the dataset. This is achieved with the definition of a hybrid cost function: \[\mathcal{F}(S,G,P,\lambda)=\lambda\frac{\sum_{i=1}^k\sum_{j>i}^k d(p_i,p_j)}{\max\limits_{p_i,p_j \in P}d(p_i,p_j)}+(1-\lambda)\frac{k(k-1)}{2}(1-rr_P(S,G))\]
where $P$ is a query result, $S \subseteq P$ is a subset of representative points, $G$ is a class of scoring functions and $\lambda$ is the parameter used to balance the tradeoff between the maximization of diversity and the minimization of the regret-ratio.

\section{Managing user preferences}

Data-driven systems have emerged as a very profitable business model during the last decade, thriving on the systematic organization of data to better serve customers' needs. Although the aspects related to modeling, integration and visualization of data definitely play a crucial role in the data-driven paradigm, it is indeed required to take into consideration the query capabilities of these systems, hence the implementation of appropriate tools that make possible the extraction of useful information from the data. As already stated in section 1, the process of information retrieval largely benefits from a user-centered approach which specifically adjust the query result to the user's preference profile, relying on a set of directly or indirectly determined input preferences. In this section we summarize the capabilities of the previously introduced query techniques to act as preference management frameworks, hence their ability to personalize the query process, control the output size, relax and adapt preference criteria.

\subsection{Preferential query answering}

The integration of user preferences in the query process is a critical aspect to take into account in many scenarios based on human-computer interaction, particularly in the web domain, where product search or recommender systems are two representative examples, but also in apparently distant domains such as cache management \cite{DBLP:conf/icde/CherniackGFZ03} or publish/subscribe systems \cite{DBLP:journals/puc/Dey01}. The first step is preference representation: this can be done in a qualitative manner, for instance using binary predicates to compare tuples, or in a quantitative manner, using scoring functions to express a degree of interest.

\begin{table}[ht]
    \centering    
    \begin{tabular}{l l c c c c}
    \hline
    \multicolumn{2}{c}{} &
    \multicolumn{2}{c}{Representation} &
    \multicolumn{2}{c}{Processing} \\
    \hline
    &
    & \raisebox{0.5 em}{\thead{Qualitative}}
    & \raisebox{0.5 em}{\thead{Quantitative}}
    & \thead{Attribute\\space}
    & \thead{Support data\\structures} \\
    \hline
      \multirow{7}{*}{\thead{FLEXIBLE\\SKYLINES}} 
      & \thead{R-Skylines \cite{DBLP:journals/pvldb/CiacciaM17}} & & \checkmark & \checkmark & \\
      & \thead{Uncertain top-k\\ queries \cite{DBLP:journals/pvldb/MouratidisT18}} & & \raisebox{0.5 em}{\checkmark} & \raisebox{0.5 em}{\checkmark} & \\
      & \thead{P-Skylines \cite{DBLP:journals/vldb/MindolinC11}} & \checkmark & & & \checkmark \\    
      & \thead{Trade-off\\ Skylines \cite{DBLP:conf/edbt/LofiGB10}} & \raisebox{0.5 em}{\checkmark} & & \raisebox{0.5 em}{\checkmark} & \raisebox{0.5 em}{\checkmark} \\
    \hline
      \multirow{4}{*}{\thead{SKYLINE\\RANKING}} 
      & \thead{Skyline\\ ordering \cite{DBLP:journals/tkde/LuJZ11}} & -- & -- & \raisebox{0.5 em}{\checkmark} & \\
      & \thead{k-Representative\\ Skylines \cite{DBLP:conf/icde/LinYZZ07}} & -- & -- & \raisebox{0.5 em}{\checkmark} & \\
      & \thead{SKYRANK \cite{DBLP:journals/dke/VlachouV10}} & \checkmark & & \checkmark & \checkmark \\    
    \hline
      \multirow{4}{*}{\thead{REGRET\\MINIMIZATION\\QUERIES}} 
      & \thead{k-Representative regret\\ minimization \cite{DBLP:journals/pvldb/NanongkaiSLLX10}} & & \raisebox{0.5 em}{\checkmark} & \raisebox{0.5 em}{\checkmark} & \\
      & \thead{k-Regret Minimizing\\ Sets \cite{DBLP:journals/pvldb/ChesterTVW14}} & & \raisebox{0.5 em}{\checkmark} & \raisebox{0.5 em}{\checkmark} & \\
      & \thead{ReDi \cite{DBLP:conf/sigmod/HussainKS15}} & & \checkmark & \checkmark & \\    
    \hline
    \end{tabular}
    \caption{Comparison of strategies used to support preferential query answering}\label{query_preferences}
\end{table}

Typically, preferences are stored in a user profile, which is then used to select, based on context information, the query preferences to adopt during the processing step. As shown in Table \ref{query_preferences}, most techniques rely on a quantitative representation of the user preferences, be it a class of scoring functions or a region surrounding a preference vector. Few other techniques instead make use of a qualitative representation using preference formulas (P-Skylines, Trade-off Skylines) or preference dimensions sets (SKYRANK).

Skyline Ranking techniques, except for SKYRANK, \textit{do not} take into consideration user query preferences, instead they rely on the properties of the skyline set, such as the maximum number of dominated points or the maximum distance between a non-representative point and its closest representative, without having a specific user in mind.

As far as the processing is concerned, practically all the examined techniques exploit the geometry of the multi-criteria optimization problem to compute the query result. It is worth highlighting few exceptions, P-Skylines, Trade-off Skylines and SKYRANK, which adopt a graph-based approach to test dominance relationships.

\subsection{Controlling output size}

The ability to formulate constraints over the query output size is crucial for the interestingness of the result, since the user might be only interested in a limited number of objects to choose from or in getting the gist of the information that resides in the dataset: if the output size is unmanageable, the decision-making process of the user can be compromised.

\begin{table}[ht]
    \centering    
    \begin{tabular}{l l c c c c}
    \hline
    & 
    & \thead{Relaxation of\\dominance}
    & \thead{User preferences\\ integration}
    & \thead{Representative\\ identification}
    & \thead{Restriction to\\ attribute subspaces} \\  
    \hline
      \multirow{7}{*}{\thead{FLEXIBLE\\SKYLINES}} 
      & \thead{R-Skylines \cite{DBLP:journals/pvldb/CiacciaM17}} & \checkmark  & \checkmark & & \\
      & \thead{Uncertain top-k\\ queries \cite{DBLP:journals/pvldb/MouratidisT18}} & \raisebox{0.5 em}{\checkmark} & \raisebox{0.5 em}{\checkmark} & & \\
      & \thead{P-Skylines \cite{DBLP:journals/vldb/MindolinC11}} & & \checkmark & & \\    
      & \thead{Trade-off\\ Skylines \cite{DBLP:conf/edbt/LofiGB10}} & \raisebox{0.5 em}{\checkmark}  & \raisebox{0.5 em}{\checkmark} & & \\
    \hline 
      \multirow{4}{*}{\thead{SKYLINE\\RANKING}} 
      & \thead{Skyline\\ ordering \cite{DBLP:journals/tkde/LuJZ11}} & & & \raisebox{0.5 em}{\checkmark} & \\
      & \thead{k-Representative\\ Skylines \cite{DBLP:conf/icde/LinYZZ07}} & & & \raisebox{0.5 em}{\checkmark} & \\
      & \thead{SKYRANK \cite{DBLP:journals/dke/VlachouV10}} & \checkmark & \checkmark & \checkmark & \checkmark \\    
    \hline
      \multirow{4}{*}{\thead{REGRET\\MINIMIZATION\\QUERIES}} 
      & \thead{k-Representative regret\\ minimization \cite{DBLP:journals/pvldb/NanongkaiSLLX10}} & & & \raisebox{0.5 em}{\checkmark} & \\
      & \thead{k-Regret Minimizing\\ Sets \cite{DBLP:journals/pvldb/ChesterTVW14}} & & & \raisebox{0.5 em}{\checkmark} & \\
      & \thead{ReDi \cite{DBLP:conf/sigmod/HussainKS15}} & & & \checkmark & \checkmark \\    
    \hline
    \end{tabular}
    \caption{Comparison of strategies used to reduce query output size}\label{output_size_control}
\end{table}

\begin{figure}[h]
\centering
\includegraphics[scale=0.7]{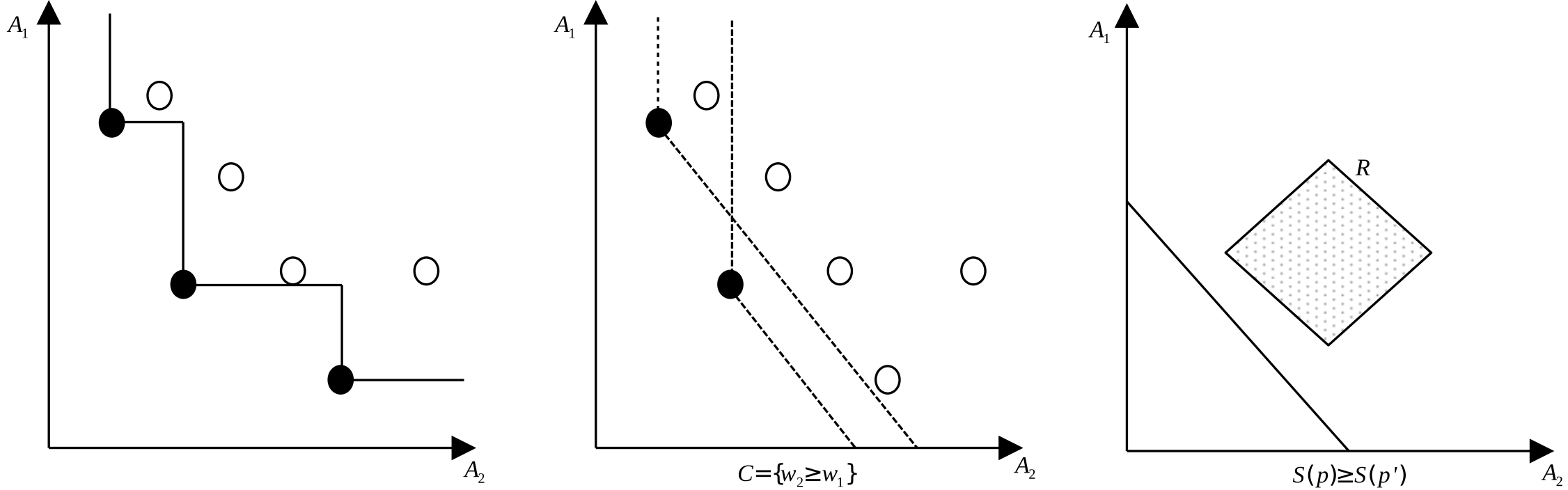}
\caption{Techniques used in Flexible Skylines to reduce query output size, from left to right: traditional Skyline, R-Skyline, Uncertain top-k queries}
\label{flexible_skylines}
\end{figure}

The reduction of the output size, as it is summarized in Table \ref{output_size_control}, can be achieved using various approaches: Skyline Ranking and Regret Minimization queries rely on the identification of representative tuples, whereas Flexible Skylines tackle the problem by relaxing the traditional definition of \textit{dominance}; the former approach aims at filtering the output set using a measure of \enquote{representativeness} which characterize the descriptive power of a tuple with respect to the full dataset, hence the use of a \textit{regret-ratio} which measures how happy the user would be if he was given this filtered output in Regret Minimization queries, or the exploitation of skyline properties to find the most dominant (hence the most likely to be interesting) tuples in Skyline Ranking: such a filtering approach allows the user to specify a parameter $k$ that represent the expected number of tuples to see as a result; the latter approach instead focuses on the extension of the dominance region (hence to the reduction of non-dominated tuples) by molding the \enquote{skyscrapers} of the skyline into \enquote{pyramids} with the use of a more general set of input preferences given by the user, or alternatively by restricting the dominance check to a specific region around the user's preference vector, as shown in Figure \ref{flexible_skylines}: in this case the result size \textit{cannot} be specified a priori, although it is worth noting that the more user's preference information is used in the query process, the more the output set is expected to decrease in size. A thorough analysis about how the problem of controlling the output size can be addressed in the context of skyline queries is proposed in \cite{DBLP:series/isrl/LofiB13}.

\subsection{Input preferences flexibility}

As previously discussed, preferential query answering is vital for finding results that best satisfy the user's requirements, yet it is a very delicate matter because of a fundamental issue, discussed in \cite{DBLP:journals/aim/PuC08}: the user is required to be the decision maker despite the exploratory nature of its query, hence it is necessary to deal with preference inconsistencies and lack of information. This particular problem is at the core of Flexible Skylines, which deal with it by overcoming the need of specifying a scoring function, thus relieving the user from the responsibility of determining exact scores for each attribute: this is achieved either by exploiting the geometry of the attribute weight space (R-Skylines, Uncertain Top-k queries) or by allowing a qualitative preference formulation (P-Skylines, Trade-off Skylines); the former method aims at generalizing the weight vector into a broader region in order to take into account possible variations of the provided weights: R-Skylines do that by asking the user a more general set of constraints that can also be more easily elicited (\textit{e.g.} price cannot be more than 3 times the mileage), whereas Uncertain Top-k queries start from a weight vector (which can be computationally inferred) and expand it into a region so that all the surrounding weight vectors are considered in the query process as well; the latter tackle the flexibility issue upstream, by using a different method not only to represent user preferences but also to extract them: P-Skylines for instance use a feedback based approach that directly or indirectly involve the user for the identification of desirable and undesirable tuples, which will be used to build its preference profile.

Skyline Ranking and Regret Minimization queries techniques on the other hand, although they also do not require the user to specify a scoring function, provide no flexibility \textit{at all}, since they only allow to specify the size parameter $k$ (or a happiness ratio parameter $\alpha$).

\section{Discussion}

In this survey we examine various techniques to take on the problem of multi-objective optimization in the context of preference-based queries. We discuss about preference representation and not only how, but also with which degree of flexibility user preferences are integrated in the query process: it emerges that a quantitative representation that makes use of scoring functions is the preferred approach, although qualitative representations are also used to take into account trade-offs or binary constraints over attributes; preferences are mostly processed directly inside the attribute space as linear constraints on attribute weights, making the dominance test a linear programming problem, despite few exceptions where a graph-based approach is used, exploiting link-based ranking techniques. We highlight the benefits on the size of the output set derived from the integration of user preference information in the query process, and we show the different control capabilities over the size parameter.

\bibliographystyle{ieeetr}
\bibliography{bibliography.bib}

\end{document}